\documentclass[%
reprint,
 amsmath,amssymb,
 aps,
]{revtex4-2}

\usepackage{subfigure}
\usepackage{float}
\usepackage{slashed}
\usepackage{xcolor}
\usepackage{amsmath}
\usepackage{subcaption}
\usepackage[english]{babel}
\usepackage{tikz}
\usepackage{adjustbox}
\usepackage{rotating}
\usepackage{rotfloat}
\usepackage{verbatim}
\usepackage[utf8]{inputenc}
\usepackage{graphicx}
\usepackage{dcolumn}
\usepackage{bm}
\usepackage[super]{nth}
\usepackage{lipsum}

\usepackage{hyperref}
\usepackage{xcolor}
\hypersetup{
  colorlinks   = true, 
  urlcolor     = red, 
  linkcolor    = blue, 
  citecolor   = blue 
}

\usepackage[compat=1.1.0]{tikz-feynman}

\usepackage{threeparttable}
\newcommand{\VLL}{L}
\newcommand{\tVLL}{\tilde{L}}

\begin{document}

\preprint{APS/123-QED}
\title{Searching for vector-like leptons decaying into an electron and missing transverse energy in e$^{+}$e$^{-}$ collisions with $\sqrt{s} = 240$ GeV at the FCC-ee}

\author{S. Elgammal}
 \altaffiliation[sherif.elgammal@bue.edu.eg]{}
\affiliation{%
Centre for Theoretical Physics, The British University in Egypt, P.O. Box 43, El Sherouk City, Cairo 11837, Egypt.
}%


\date{\today}

\begin{abstract}
{This analysis delves into the vector-like lepton model decaying to lepton and dark matter by utilizing samples based on Monte Carlo simulations from $e^+ e^-$ collisions at the Future Circular Collider (FCC-ee). It will operate at a center-of-mass energy of 240 GeV and an integrated luminosity of 10.8 ab$^{-1}$. 
The key signal signature features missing transverse energy alongside dilepton events. Should these new particles not be detected, this study establishes 95\% confidence level exclusion limits on the vector-like lepton mass and the Yukawa coupling.}

\end{abstract}

\maketitle

\section{Introduction}
\label{sec:intro}
The Standard Model (SM) of particle physics offers an efficient framework for describing well-known elementary particles alongside their interactions. However, it does not include a viable candidate for dark matter (DM), prompting researchers to investigate extensions beyond the Standard Model (BSM). 
These extensions aim to clarify the nature of DM and its possible interactions with visible matter.
The weakly interacting massive particles (WIMPs)~\cite{wimps} are the most extensively studied candidates for DM in these new models. 
These particles could account for recent astrophysical observations \cite{planck2015,planck2018}, which indicate that DM constitutes approximately 25\% of the universe's energy, while visible matter makes up only about 5\%.

The lepton-portal dark matter model, initially proposed in references \cite{Chang:2014tea, Bai:2014osa}, features a singlet DM particle that interacts with the SM group through leptonic interactions. In this framework, dark matter annihilates into SM leptons via the t-channel exchange of vector-like leptons (VLLs). A notable feature of this model is that these VLLs have both left- and right-handed components, which transform identically under the SM electroweak (EW) gauge symmetry.

VLLs are a common prediction in various frameworks that extend BSM, including scenarios like supersymmetry \cite{susy1, susy2, susy3}, grand unified theories \cite{gut1, gut2, gut3}, and models that incorporate extra spatial dimensions \cite{extra_dim1, extra_dim2}. They are often referenced in DM theories \cite{vll_dm1, vll_dm2, vll_dm3, vll_dm4}, and can shed light on the observed hierarchies in fermion masses across different generations \cite{hierarchy1, hierarchy2, hierarchy3}. Additionally, VLLs may offer a solution to the gap between the SM's predictions and recent experimental observations regarding the muon anomalous magnetic moment \cite{susy3, muon_moment1, muon_moment2, muon_moment3, muon_moment4, lpdm, Kawamura:2022uft}.

The CMS collaboration has analyzed vector-like tau leptons in hadronic and leptonic final states \cite{cms_vll}, establishing cross-section limits for both singlet and doublet models. In the doublet model, vector-like leptons coupled to third-generation SM leptons are excluded for masses from 100 up to 1040 GeV, with an expected limit from 100 to 970 GeV. For VLL masses below 280 GeV, the most stringent limits come from a model-independent scheme, while model-dependent BDT regions apply for masses above this threshold. In the singlet model, the strongest limits apply across the mass range, with observed exclusions between 125–170 GeV and an expected exclusion at approximately 150 GeV.

To complement the results from CMS, the ATLAS collaboration searched for vector-like leptons, either as doublets or singlets, which couple to the first and second generation of SM leptons \cite{vll_atlas}. The data aligned well with SM predictions, establishing exclusion limits on the masses of vector-like electrons and muons. Specifically, masses were excluded in the range of 100 to 1220 GeV for vector-like electrons and 100 to 320 GeV for vector-like muons, with the ranges extending to 1270 GeV and 400 GeV, respectively, in the doublet and singlet scenarios, all at a 95\% confidence level.

The current study investigates a specific type of very light particle, referred to as VLL, with masses below the limits set by the CMS and ATLAS collaborations (specifically, $M_{VLL} < 100$ GeV). This particle can decay into SM electron, with the resulting missing transverse energy ($E_{T}^{miss}$) being linked to dark matter (DM). We explore one scenario for the mass difference between the VLL and the scalar dark matter ($\Delta M$): $\Delta M = 5$ GeV. Details of the production mechanism are discussed in Section \ref{section:model}.

We aim to explore the production of VLL pairs in $e^+ e^-$ collisions at the Future Circular Collider (FCC-ee) \cite{FCC}, utilizing this opportunity to delve into the properties of DM. The FCC-ee, a foreseen $e^+ e^-$ collider at CERN, is set to operate with a center-of-mass energy ($\sqrt{s}$) ranging from 240 GeV to 365 GeV. Though other future lepton colliders, such as CLIC \cite{CLIC}, are also under consideration, our study will concentrate primarily on the physics possibilities offered by the FCC-ee.


This paper is structured as follows. First, in Section \ref{section:model}, we provide a concise introduction to the theoretical framework underlying the VLL model. Section \ref{section:MCandDat} discusses the simulation methods used to create both the model signal and SM background samples. The criteria for event selection and the analysis strategy are outlined in Section \ref{section:AnSelection}. Finally, we present our results in Section \ref{section:Results}, followed by conclusions and a summary of the analysis in Section \ref{section:Summary}.

\section{The simplified VLL model}
\label{section:model}
\begin{figure}
\centering
\subfigure[]{
  \includegraphics[width=60mm]{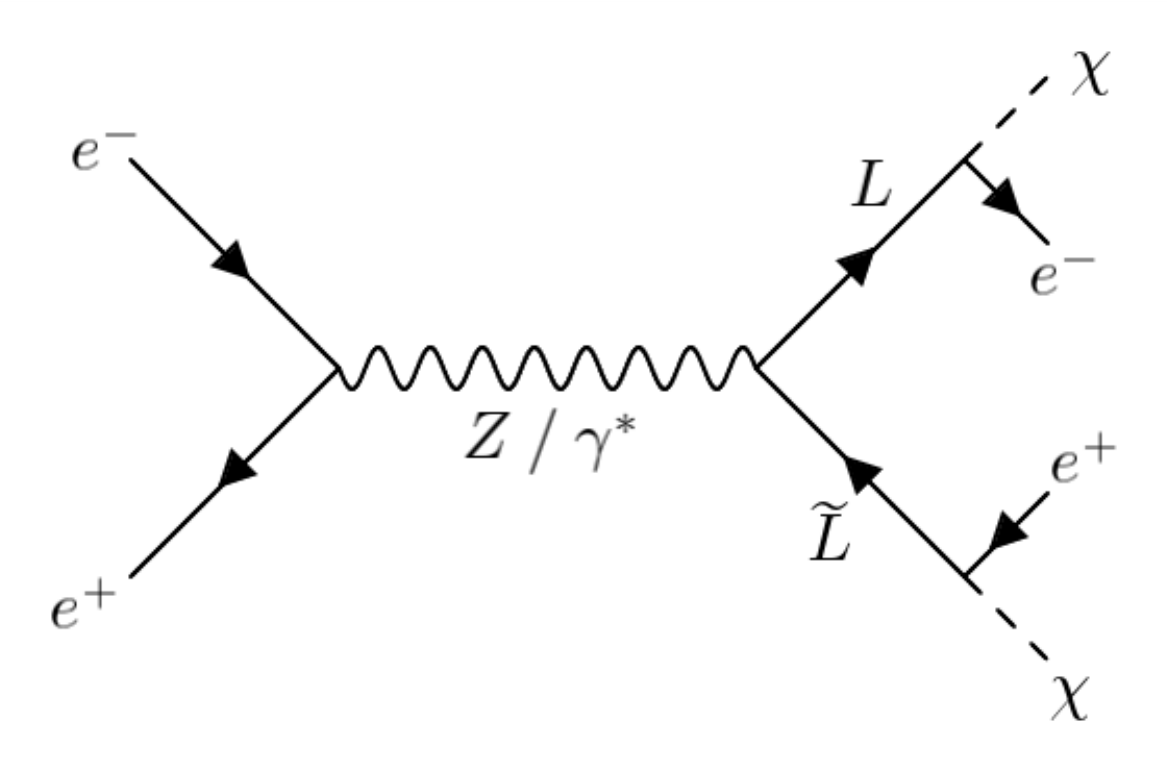}
  \label{fey1}
}
\subfigure[]{
  \includegraphics[width=40mm]{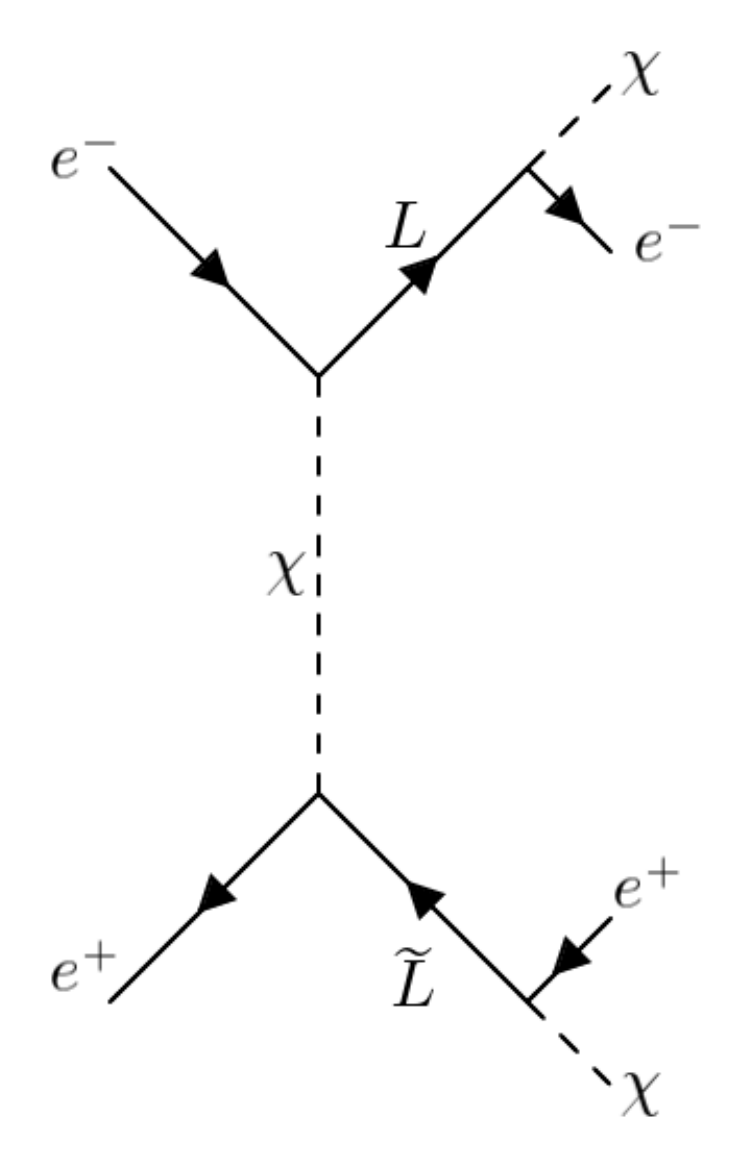}
  \label{fey2}
}
   \caption{Feynman diagrams illustrating the production of VLL pairs ($L\bar{L}$) from the s-channel \ref{fey1} and t-channel \ref{fey2}, which are then followed by the decay of $L$ to electron and dark matter ($\chi$).}
   \label{figure:fig1}
\end{figure}
We will examine the VLL model, where the DM particle ($\chi$) is a singlet under the SM's gauge group. Thus, the DM can either be a scalar or a fermion. In this scenario, the mediator is a VLL. Since $\chi$ is an SM singlet, it interacts with the SM via Yukawa couplings ($\lambda^{i}_{L}$) involving additional leptons that possess the gauge quantum numbers of $(2, -1/2)$ under the electroweak gauge group $SU(2)\times U(1)$. 
To get rid of any anomalies, additional leptons have been designed to be vector-like. We refer to the electroweak doublet vector-like leptons as $\VLL$. Moreover, we have imposed a $Z_2$ symmetry or a global $U(1)$ symmetry to guarantee the stability of DM. Under this added symmetry, all SM fields transform straightforwardly, while the DM and the extra leptons experience more complex transformations.
The complete details of this model can be found in \cite{lpdm}. 

The Lagrangian that contains the VLL mass and Yukawa interaction terms is introduced by 
\begin{align}
  -\mathcal{L} 
  = M_L \bar{L}_L L_R 
          +\lambda_L^i X\bar{\ell}_{L_i} L_R 
          .
\end{align}

These VLLs are thought to couple exclusively with either the first or second generation of SM leptons. This limitation is put in place to prevent lepton flavor violations, which experimental results have strongly constrained. In this analysis, we concentrate on the doublet vector-like lepton that interacts with the electron. Two key parameters are particularly important for this study: the mass of the VLL, denoted as \( M_L \), and the Yukawa coupling \( \lambda_L := \lambda_L^1 \). 


In our study, we explore the generation of pairs of VLLs and their subsequent decay into an SM electron alongside a scalar DM particle. 
These VLLs can be generated through an s-channel process, as shown in Figure \ref{fey1}, where interactions occur via $Z/\gamma^*$. Additionally, they can also emerge from a t-channel process, as illustrated in Figure \ref{fey2}, in which the scalar DM particle $\chi$ acts as a mediator. 
The t-channel process contribution is heavily influenced by the strength of the coupling constant $\lambda_L$. When $\lambda_L$ is low, this contribution tends to be negligible.
The exact signature of this process is marked by the presence of two electrons and a notable amount of $E_{T}^{miss}$, which can be linked to stable DM particles.

The model has several free parameters, including the mass of DM $(M_{\chi})$, the mass of the VLLs $(M_{L})$, and the Yukawa coupling $(\lambda_{L})$. Additionally, the mass difference between the VLL and the DM is denoted as $\Delta M$.

\section{Simulation of the VLL signal samples and the SM backgrounds and systematic uncertainties}
\label{section:MCandDat}
For the SM background processes that can generate electron pairs in the signal region, we take into account several important contributions. These primarily involve decays that produce pairs of electrons ($Z/\gamma^* \rightarrow e^+e^-$) and tau leptons ($Z/\gamma^* \rightarrow \tau^+\tau^-$).
Also, we have the production of top quark pairs, which can manifest as $\text{t}\bar{\text{t}} \rightarrow e^+e^- + 2b + 2\nu$, along with diboson production. This encompasses processes such as $W^{+}W^{-} \rightarrow e^+e^- + 2\nu$, $ZZ \rightarrow e^+e^- + 2\nu$ and $ZZ \rightarrow 4e$.

\subsection{Monte Carlo simulation}
\label{systematic}

\begin{table}[h!]
\small
\centering
\caption{The MC simulated SM samples are produced from $e^+ e^-$ collisions at the FCC-ee, specifically at $\sqrt{s} = 240$ GeV. This includes their cross-sections multiplied by branching ratios, along with their order of generation. Additionally, the names of each sample and the generator package utilized are also provided.}
\begin {tabular} {|c|c|c|c|c|}
\hline
Process \hspace{0.5cm} & Deacy channel  & Generator  & {$\sigma \times \text{BR} ~(\text{fb})$} & Order \\
\hline
\hline
$Z/\gamma^{*}$ & $e^{+}e^{-}$ & Whizard & 2173.0 & LO\hspace{6cm}\\
\hline
$Z/\gamma^{*}$ & $\tau^{+}\tau^{-}$ & Whizard & 2231.2 & LO\hspace{6cm}\\
\hline
WW & $e^{+}e^{-} + 2\nu$ & Whizard & 202.1 & LO \\
\hline
ZZ & $e^{+}e^{-} + 2\nu$  & Whizard & 5.0 & LO \\
\hline
ZZ  & $4e$ & Whizard & 0.6 & LO \\
\hline
$\text{t}\bar{\text{t}}$ & $e^{+}e^{-} + 2\nu + 2b$ & Whizard& 1.7$\times 10^{-6}$ & LO \\
\hline
\end {tabular}
\vspace{4pt}
\label{table:tab3}
\end{table}

The VLL model signal samples and the associated SM background samples have been produced privately. 
They have been produced via the WHIZARD generator package version 3.1.1 \cite{whizard,Omega}. The impact of initial state radiation (ISR) has been considered and was convoluted with Pythia 6.24 to handle the hadronization and the modeling of the parton shower \cite{pythia6}. 
For a fast simulation of the IDEA detector model \cite{idea}, we employed the DELPHES package \cite{delphes}.
These simulations have been based on $e^+ e^-$ collisions conducted at the FCC-ee with $\sqrt{s} =$ 240 GeV, reflecting the conditions expected during RUN 1.

\begin{figure} [h!]
\centering
\resizebox*{9.cm}{!}{\includegraphics{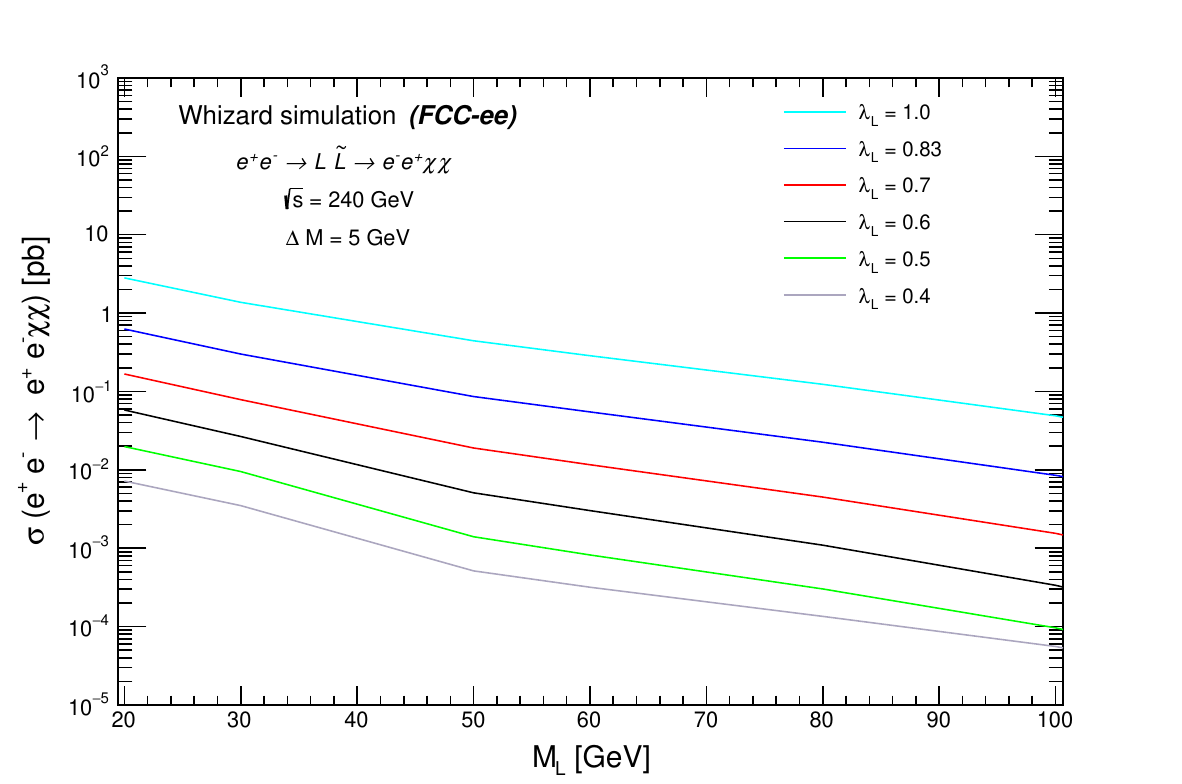}}
\caption{The dependence of cross sections for the signal process \( e^{+} e^{-} \rightarrow L\tilde{L} \rightarrow e^{+}e^{-} \chi\chi \), as induced by the VLL model, on the VLL mass is analyzed at \( \sqrt{s} = 240 \, \text{GeV} \). This analysis considers various values of \( \lambda_{L} \) and a mass difference \( \Delta M = 5 \, \text{GeV} \).}
\label{figure:SigmaVsMd}
\end{figure}

We have produced signal events by generating pairs of VLLs that decayed into DM and SM electrons. The decay process adhered to the following formula \cite{lpdm}
\begin{align}
e^+ e^- \rightarrow \VLL {\tVLL},~~\VLL 
\rightarrow e^{-} \chi,~~{\tVLL} \rightarrow e^{+} \chi.
\end{align}

The research explored various scenarios involving mass differences between VLLs and DM, specifically examining cases where the mass-splitting, $\Delta M$, is set at 5 GeV.

The production of VLLs has been examined through t-channel processes driven by DM, alongside s-channel production methods. Notably, the VLL (\(\VLL\)) decays exclusively into an electron and the DM particle (\(\chi\)), resulting in a branching ratio of one. 
The mass of \(\chi\) is vital in influencing the contributions from the t-channel process. When the Yukawa couplings are large enough, the mass of DM has a significant impact on the cross-section values.

We carried out multiple sets of event generation, beginning with a lower Yukawa coupling of \(\lambda_L = 0.4\) and scaling up to \(\lambda_L = 1\). With the smaller Yukawa coupling, the contribution from the t-channel process remained negligible. In contrast, as we explored the higher Yukawa coupling, we noticed a considerable increase in the contribution from the t-channel process. 

The production cross-sections were plotted against $M_{L}$ for a range of \(\lambda_{L}\) values, as illustrated in Figure \ref{figure:SigmaVsMd}, using a small mass difference scenario (\(\Delta M = 5\) GeV). The results reveal that in scenarios featuring a larger Yukawa coupling, the production cross-section is notably higher compared to cases with a smaller Yukawa coupling, underscoring the dominance of the t-channel process.

Monte Carlo simulations were used for generating SM background samples and calculating their cross-sections, as shown in Table \ref{table:tab3}. The very small cross-section for the leptonic decay of \( t\bar{t} \) led to its exclusion from the analysis.

Utilizing these simulations, we estimated both the signal samples and the SM background processes, normalizing them to their respective cross-sections and an integrated luminosity ($\mathcal{L_{\text{int}}}$) of 10.8 ab$^{-1}$.

\subsection{Systematic uncertainties}
\label{systematic}
To account for possible systematic effects, we applied a flat uncertainty of 5\% uniformly across the board, distributed as follows. In the context of circular colliders, the contribution of beamstrahlung to the uncertainty in center-of-mass energy is significantly lower, around 1\% \cite{idea}. This allows for tighter constraints on the beam energy. A notable scenario where this is particularly beneficial is in determining the Higgs recoil mass. 

With respect to the beam energy spread (BES), it could achieve a measurement precision better than 1\% by utilizing the abundant events from $e^{+}e^{-} \rightarrow \mu^{+}\mu^{-}(\gamma)$ and $e^{+}e^{-} \rightarrow e^{+}e^{-}(\gamma)$ radiative returns within the same dataset \cite{delphes-systematics}. 

For the IDEA detector concept for FCC-ee, the electromagnetic energy resolution can be effectively characterized as a function of the electron energy, ranging from 0.6\% to 5\% as the electron energy decreases. Finally, the transverse momentum resolution stands at 0.3\% for a momentum of 100 GeV/c \cite{idea}.

\section{Event selection and background reduction}
\label{section:AnSelection}

\subsection{The pre-selection of event}
\label{section:finalcuts}
The event has been initially selected to reconstruct a final state that features two opposite-charge electrons with low transverse momentum $(p^{e}_{T})$ and incorporates $E_{T}^{miss}$ to account for invisible particles (DM). Various kinematic parameters are carefully examined, with specific cuts applied to refine the selection.

Both electrons must meet a set of preliminary selection criteria. These include $p^{e}_{T}$ greater than 5 GeV, a pseudo-rapidity ($|\eta^{e}|$) less than 2.5, and the ratio of energy deposited in the hadronic calorimeter to that in the electromagnetic calorimeter ($E_{\text{had}}/E_{\text{em}}$) should be under 0.1.
Specifically, this ratio, indicating the proportion of energy in the hadronic versus electromagnetic calorimeters, should remain below 10\%. The pre-selection requirements are detailed in Table \ref{cuts-pre}.

\begin{table} [h!]
  \vspace{3pt}
    \caption{The cut-based event pre-selection.}
    \centering
    \begin{tabular}{|c|c|}
\hline
variable & optimized cut value  \\
\hline
    \hline
  $p^{e}_{T}$ (GeV)   &  $>$ 5    \\
      &  \\
  $|\eta^{e}|$    & $<$ 2.5 \\
      &  \\
  $E_{\text{had}}/E_{\text{em}}$ & $<$ 0.1  \\
      \hline
    \end{tabular}
    \label{cuts-pre}
\end{table}

The physics observable $H_{P} + E_{T}^{miss}$ is utilized to estimate the sensitivity of the VLL model \cite{cms_vll, vll_atlas}. In this context, $H_{P}$ represents the scalar sum of the transverse momenta of all charged electrons involved in the channel.

\begin{figure} 
\centering
  \resizebox*{8.0cm}{!}{\includegraphics{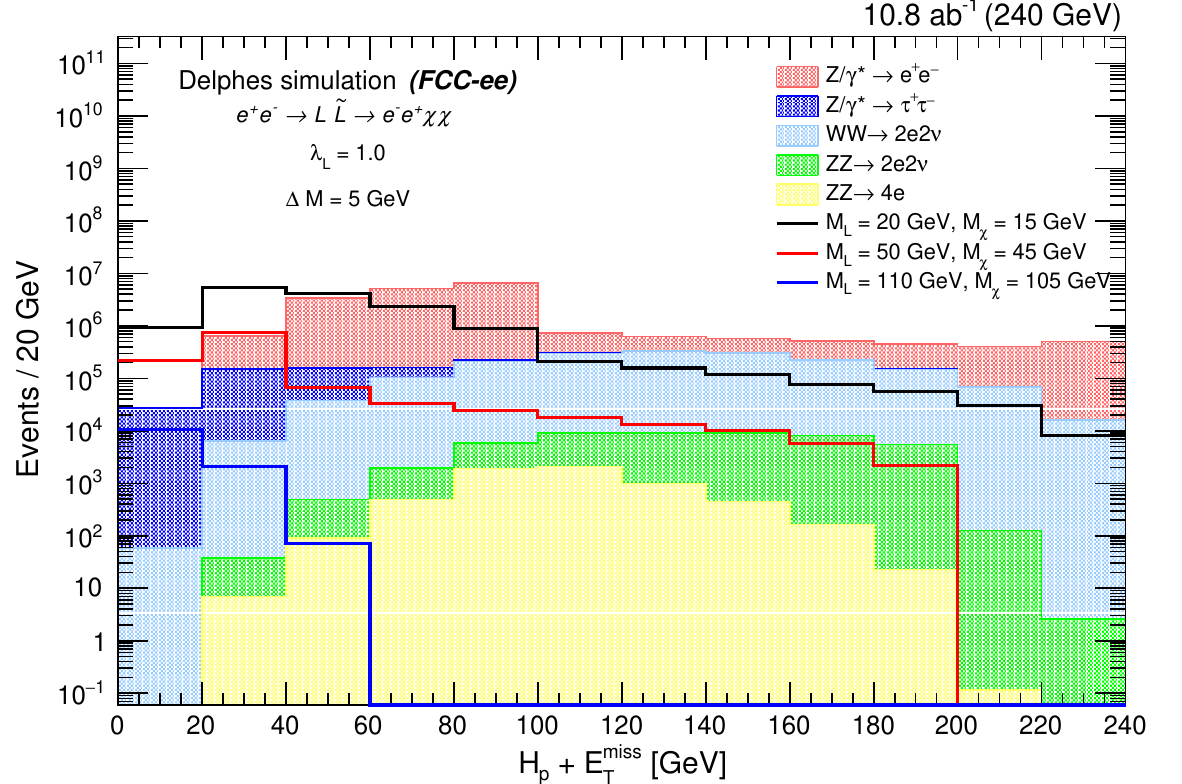}}
  \caption{The $H_{P} + E_{T}^{miss}$ distributions are presented after applying the pre-selection criteria listed in Table \ref{cuts-pre}. This spectrum includes estimates of the SM backgrounds alongside various vector-like lepton (VLL) masses generated according to the VLL model. The cases examined correspond to $\lambda_{L} = 1.0$ and mass differences $\Delta M =$ 5 GeV.}
\label{figure:fig3}
\end{figure}

Figure \ref{figure:fig3} presents the distribution of $H_{P} + E_{T}^{miss}$ for events that met the pre-selection cuts detailed in Table \ref{cuts-pre}. Such that the red histogram illustrates the background from $Z/\gamma^{*} \rightarrow e^+e^-$ events, and the blue histogram depicts the background from $Z/\gamma^{*} \rightarrow \tau^+ \tau^-$ processes. The cyan histogram accounts for the di-boson background ($WW$), and the green histogram represents the $ZZ\rightarrow 2e2\nu$ process. Additionally, the yellow histogram showcases the events from $ZZ\rightarrow 4e$. 
These histograms have been plotted in a stacked format for clarity. The signals from the VLL model, generated with varying masses ($M_{L}$= 20, 50, and 110 GeV) while keeping the coupling constants fixed at $\lambda_{L} = 1.0$ and $\Delta M = 5$ GeV, are represented by different colored lines superimposed on the histograms.

This figure illustrates that the signal events are considerably blended within background events throughout the entire $H_{P} + E_{T}^{miss}$ spectrum for \(\Delta M = 5\) GeV. 
Consequently, as will be discussed in the following paragraph, it is necessary to implement stronger cuts to distinguish the signal samples from the SM backgrounds.

\subsection{The final selection of event}
\label{section:finalcuts}


\begin{figure*}
\centering
\subfigure[]{
  \includegraphics[width=80mm]{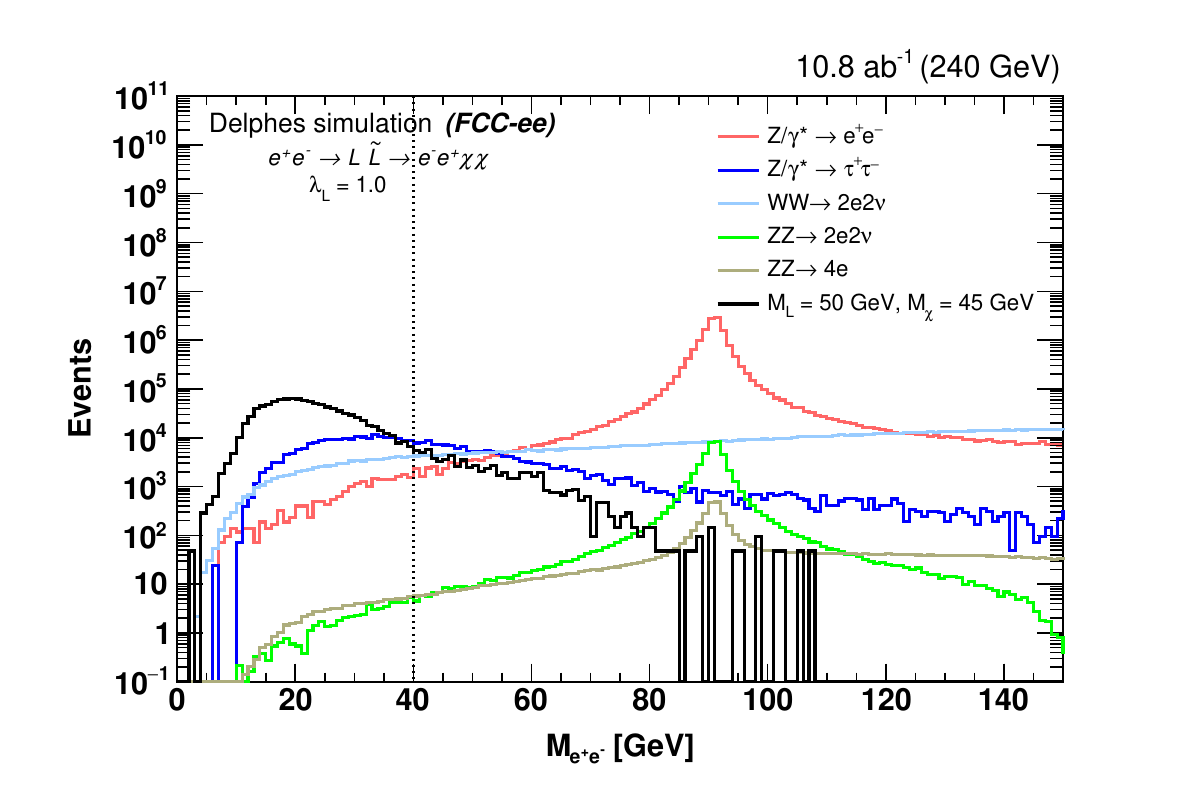}
  \label{figure:mass}
}
\subfigure[]{
  \includegraphics[width=80mm]{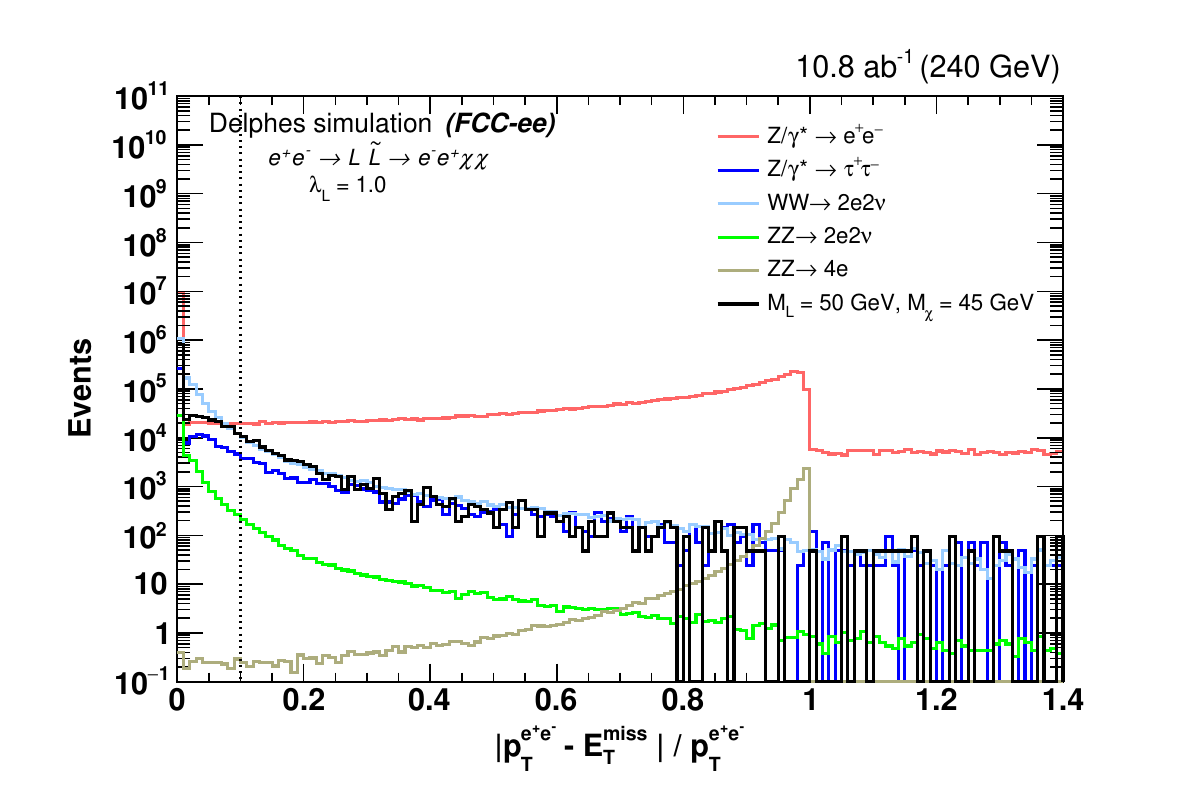}
  \label{figure:ptdiff}
}
\\
  \subfigure[]{
  \includegraphics[width=80mm]{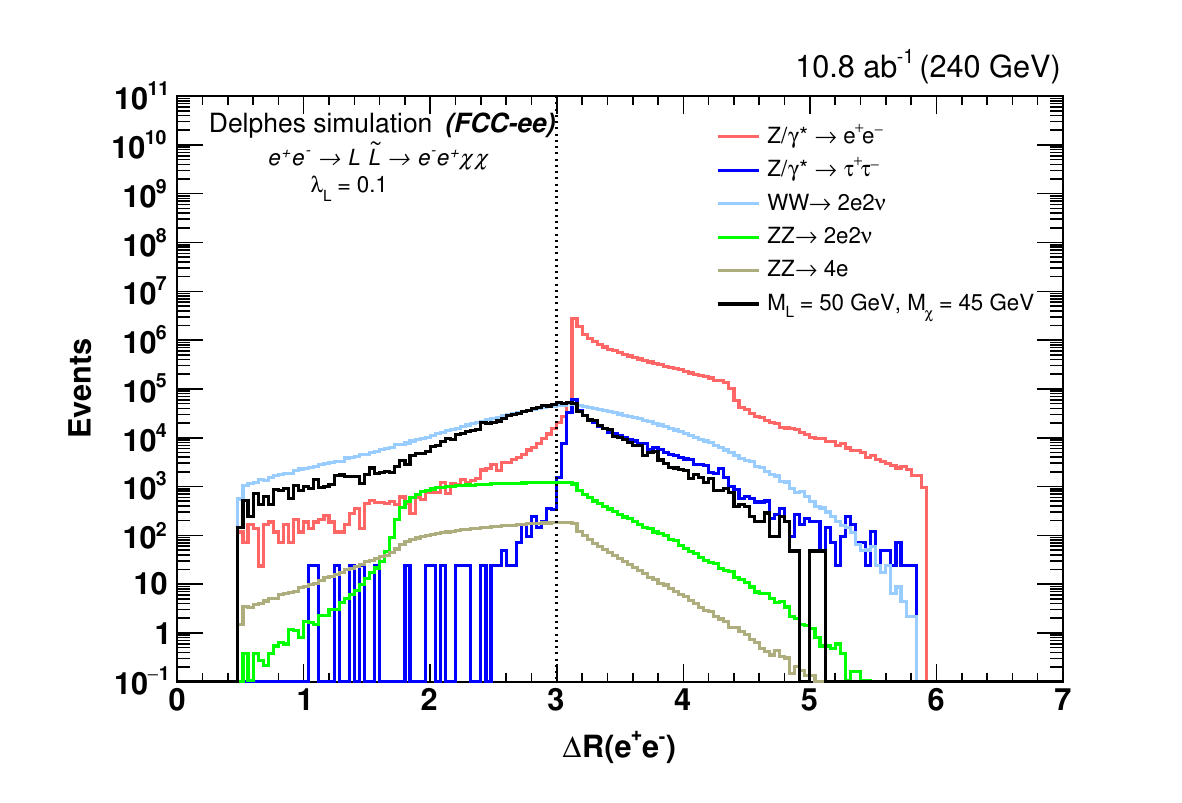}
  \label{figure:deltar}
}
\subfigure[]{
  \includegraphics[width=73mm]{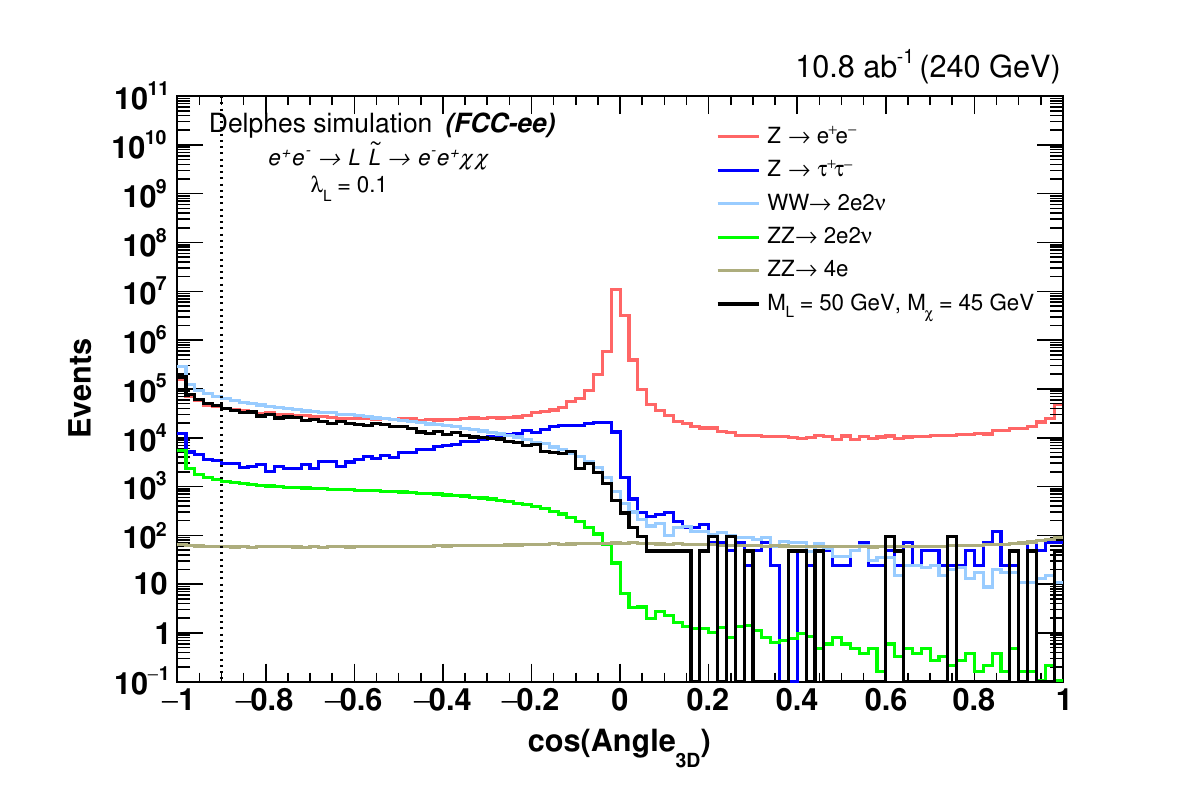}
  \label{figure:3D}
}
\caption{The four variables distributions for dielectron events, with each electron passing the pre-selection cuts listed in Table \ref{cuts-pre}. 
The variables are $M_{e^+e^-}$ \ref{figure:mass},
$|p_{T}^{e^{+}e^{-}} - E_{T}^{miss}|/p_{T}^{e^{+}e^{-}}$ \ref{figure:ptdiff},
$\Delta R(e^{+}e^{-})$ \ref{figure:deltar}, and
$\text{cos}(\text{Angle}_{3D})$ \ref{figure:3D}.
The model demonstrates the VLL with $M_{L} = 50$ GeV and $M_{\chi} = 45$ GeV, and for the SM background samples. The black vertical dotted lines represent the chosen cut value.}
\label{figure:fig70}
\end{figure*}

To refine our selection process, we implemented stricter criteria based on four essential variables. Firstly, we set a limit on the di-electron invariant mass ($M_{e^+e^-}$), requiring it to be less than 40 GeV. Secondly, we assessed the relative difference between the transverse momentum of the di-electron pair \( (p_{T}^{e^{+}e^{-}}) \) and the missing transverse energy \( E_{T}^{\text{miss}} \). This was defined by the condition that the absolute difference \( |p_{T}^{e^{+}e^{-}} - E_{T}^{\text{miss}}|/p_{T}^{e^{+}e^{-}} \) must be less than 0.1. 
Next, we stipulated that the angular separation \( \Delta R(e^{+}e^{-}) \) between the two opposite-sign electrons should be less than 3.0.
Lastly, we introduced a criterion concerning the cosine of the 3D angle ($\text{cos}(\text{Angle}_{3D})$) between di-electron system vector and the missing energy vector, ensuring they are aligned back-to-back, with the requirement that \( \cos(\text{Angle}_{3D}) \) be less than -0.9. 
The list of this tight selection is presented in Table \ref{cuts}.

\begin{table} [h!]
\caption{The cut-based event final selection.}
    \centering
    \begin{tabular}{|c|c|}
\hline
Step & Cuts optimized value  \\
\hline
    \hline
                  &     \\
   Pre-selection  & Table \ref{cuts-pre}  \\
                  &   \\
\hline
&$M_{e^{+}e^{-}} <$ 40 GeV \\
Tight selection&$|p_{T}^{e^{+}e^{-}} - E_{T}^{\text{miss}}|/p_{T}^{e^{+}e^{-}} <$ 0.1 \\
&$\Delta R(e^{+}e^{-}) < 3.0$\\
&$\text{cos}(\text{Angle}_{3D})< -0.9$ \\
    \hline
    \end{tabular}
    \label{cuts}
\end{table}

The histograms displayed in Figure \ref{figure:fig70} showcase the distributions of these variables for a single signal presentation of the VLL model, featuring \(M_{L}\) value of 50 GeV, DM mass of \(M_{\chi} = 45\) GeV, and $\lambda_{L} = 1.0$. This figure also includes the SM backgrounds. 
These variables are depicted for both the VLL signal sample and the SM backgrounds, specifically in relation to dielectron events that adhere to the pre-selection cuts detailed in Table \ref{cuts-pre}.

We begin by examining the di-electron invariant mass, denoted as \( M_{e^{+}e^{-}} \), which can be found in Figure \ref{figure:mass}. Following that, we look at the second variable: the relative difference calculated as \( |p_{T}^{e^{+}e^{-}} - E_{T}^{\text{miss}}|/p_{T}^{e^{+}e^{-}} \), illustrated in Figure \ref{figure:ptdiff}. Next, we present \( \Delta R(e^{+}e^{-}) \) in Figure \ref{figure:deltar}. Finally, we display the cosine of the three-dimensional angle, represented as \( \text{cos}(\text{Angle}_{3D}) \) in Figure \ref{figure:3D}. In each of these figures, dotted vertical black lines indicate the selected cut values.
\section{Results}
\label{section:Results}
\begin{figure} 
\centering
  \resizebox*{8.0cm}{!}{\includegraphics{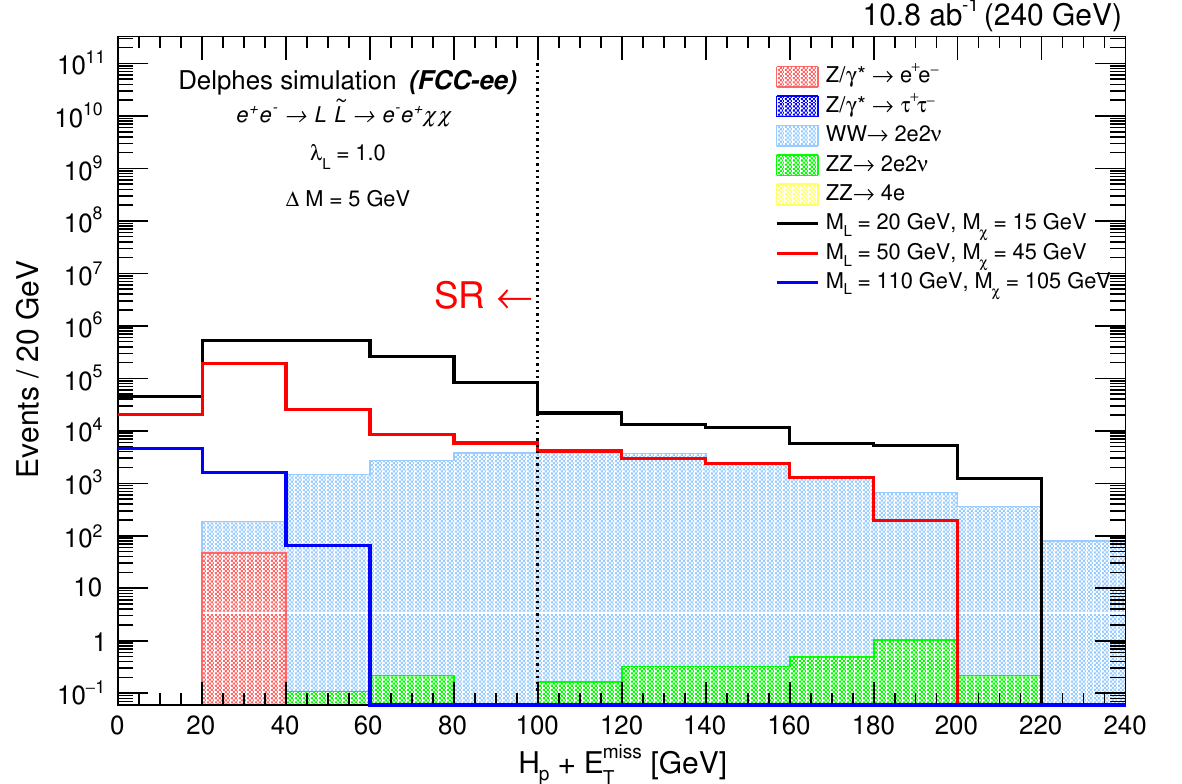}}
  \caption{The $H_{P} + E_{T}^{miss}$ distributions, for events passing the final-selection cuts listed in table \ref{cuts}, for the SM backgrounds and different values of $M_{L}$ produced in the framework of the VLL model, with fixed parameters $\lambda_{L} = 1.0$, and $\Delta M = 5$ GeV.}
\label{figure:missfinal}
\end{figure}

The analysis's results have been obtained through a shape-based analysis, with the variable $H_{P} + E_{T}^{miss}$ acting as the primary differentiator. Figure \ref{figure:missfinal} displays the distributions of $H_{P} + E_{T}^{miss}$ for both signal and SM background samples, focusing on VLLs with mass differences $\Delta M = 5$. 
These distributions are related to the electro-philic scenario with $\lambda_L = 1.0$, produced from $e^+ e^-$ collisions at the FCC-ee with $\sqrt{s} = 240$ GeV and $\mathcal{L_{\text{int}}} = 10.8$ ab$^{-1}$. Event selection is carried out based on the final cuts detailed in Table \ref{cuts}.

\begin{table*}
\fontsize{8.pt}{11pt}
\caption{The number of events that passed the pre-selection (middle column) and final-selection (right column) cuts for signal and SM background samples, performed at $\mathcal{L_{\text{int}}} = 10.8$ ab$^{-1}$ and $\sqrt{s} = 240$ GeV. The signals represent the VLL model with $\Delta M$ = 5 GeV, $\lambda_{L} = 1.0$. The total uncertainties include statistical and systematic errors.}
\small
    \centering
    
    \begin{tabular}{|c|c|c|}
\hline
Process & No. of events passing pre-selection& No. of events passing final selection \\
\hline
\hline
$Z/\gamma^{*} \rightarrow e^+ e^-$ & $17487774.7 \pm 874398.7$  & 
$ 46.9 \pm   7.2$\\
$Z/\gamma^{*} \rightarrow \tau^+ \tau^-$ & $392856.0 pm 19652.8$ & 
$0.0 \pm  0.0$\\
$\text{WW} \rightarrow e^{+}e^{-} + 2\nu$ & $1697659.9 \pm 84893.0$ &
$20499.9 \pm 1034.9$ \\
$\text{ZZ} \rightarrow e^{+}e^{-} + 2\nu$ & $43063.7 \pm 2163.2$ & 
$  2.9 \pm   1.7$  \\
$\text{ZZ} \rightarrow 4e$ & $6254.1005 \pm 322.5500$ & 
$  0.0 \pm   0.0$ \\
Sum Bkgs & $19627608.7 \pm 981390.4$ & 
$20549.7 \pm 1037.4$  \\
\hline
Signal of VLL model  &&  \\
(at $M_{L}$ = 20 GeV and $M_{\chi} = 15$ GeV) & $14437028.8 \pm 721861.4$ & 
$1502155.3 \pm 75117.8$ \\

(at $M_{L}$ = 50 GeV and $M_{\chi} = 45$ GeV) & $1126939.4 \pm 56357.0$ & 
$266082.2 \pm 13314.1$  \\

(at $M_{L}$ = 110 GeV and $M_{\chi} = 105$ GeV) & $12618.9 \pm 640.9$ & 
$6347.5 \pm 327.2$ \\
\hline
\end {tabular}
\label{table:tab18}
\end{table*}

Table \ref{table:tab18} presents the count of events that successfully fulfilled the pre-selection cuts (illustrated in the middle column) and the complete final-selection cuts (shown in the right column). 
These findings stem from simulations that took into account various SM backgrounds as well as three signal samples from the VLL model, detailed in the left column. The simulations were conducted with $\mathcal{L_{\text{int}}} = 10.8$ ab$^{-1}$ at a $\sqrt{s} = 240$ GeV. 
The signal samples are presented for the VLL model with $M_{L} =$ 20, 50, and 110 GeV, and the following fixed parameters $\Delta M = 5$ GeV and $\lambda_{L} = 1$. We accounted for total uncertainties that encompass both statistical and systematic errors. 

The results presented in Table \ref{table:tab18} clearly illustrate that the final selection cuts effectively eliminate events from \(Z/\gamma^{*}\rightarrow \tau\tau\) and \(ZZ\rightarrow4e\). Moreover, we see a notable reduction in events from \(Z/\gamma^{*}\rightarrow e^{+}e^{-}\), \(ZZ\rightarrow2e2\nu\), and WW. It is worth mentioning that some signal samples face a drop of over 75\% due to these selection criteria. Nevertheless, the overall signal shape continues to be distinguishable from the SM backgrounds within the signal region (SR), particularly at lower values of $H_{P} + E_{T}^{miss}$ (i.e., $H_{P} + E_{T}^{miss} <$ 100 GeV).

We used the profile likelihood method to analyze our results, and we conducted a statistical test based on $H_{P} + E_{T}^{miss}$ distributions. To obtain the exclusion limits on the product of signal cross sections at a 95\% CL, we employed the modified frequentist construction CLs \cite{R58, R59}, which relies on the asymptotic approximation \cite{R2}. In this likelihood framework, we treated the systematic uncertainties as nuisance parameters.

\begin{figure} 
\centering
  \resizebox*{8.0cm}{!}{\includegraphics{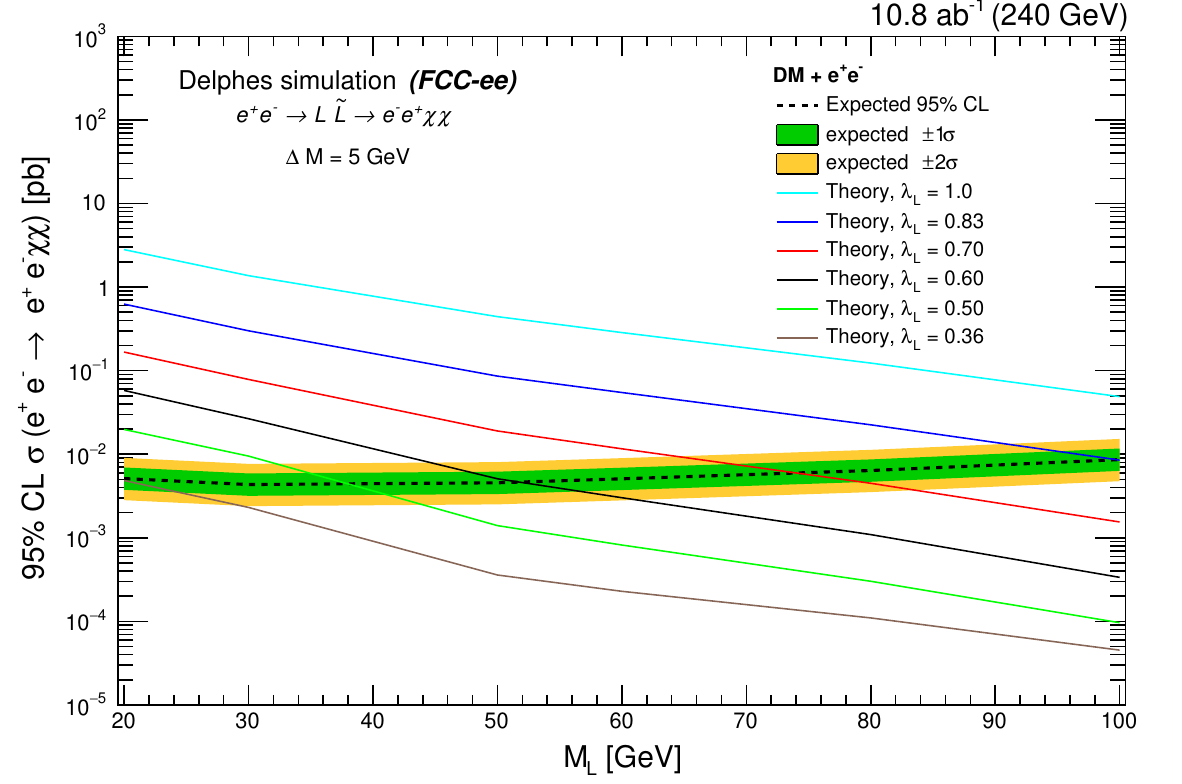}}
  \caption{The 95\% CL limit has been established on the expected cross-section of VLL
against $M_{L}$. The solid colored lines represent the cross-sections predicted by the electro-philic VLL model for several values of $\lambda_{L}$, specifically for mass differences of $\Delta M = 5$ GeV.}
  \label{figure:fig7}
\end{figure}

Figure \ref{figure:fig7} displays the 95\% CL limits on the cross-section for the VLL simplified model, specifically focusing on the electronic decay of $\VLL$. The results are derived from small mass differences of $\Delta M = 5$ GeV, based on $\mathcal{L_{\text{int}}} = 10.8$ ab$^{-1}$ and $\sqrt{s} = 240$ GeV. 
Distinct solid lines in various colors represent these limits, each corresponding to different values of the coupling constant $\lambda_{L}$. Furthermore, an expected limit, indicated by a dotted black line, arises from the statistical test; meanwhile, the yellow and green bands highlight the $\pm2$ and $\pm1$ $\sigma$ ranges, respectively.

\begin{figure} 
\centering
  \resizebox*{8.0cm}{!}{\includegraphics{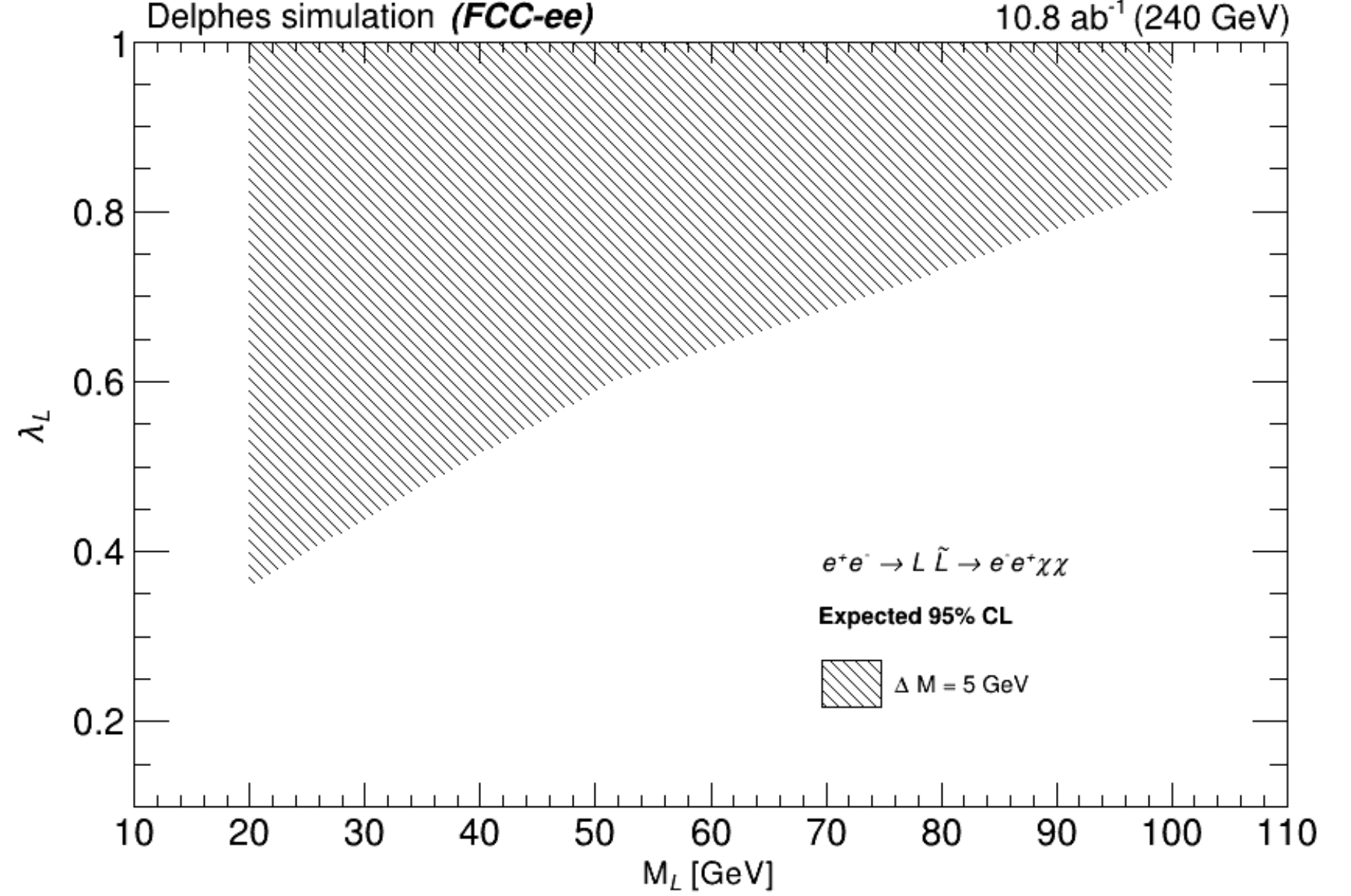}}
  \caption{The excluded parameter space region at 95\% CL in the $M_{L}$ and $\lambda_{L}$ plane for the VLL model at $\sqrt{s} = 240$ GeV and $\mathcal{L_{\text{int}}} = 10.8$ ab$^{-1}$. The contour represents the excluded region for a given value of $\Delta M =$ 5 GeV.}
  \label{figure:fig8}
\end{figure}

The results for the VLL model are interpreted within the $M_{L}-\lambda_{L}$ plane, considering $\Delta M =$ 5 GeV. The results presented in Figure \ref{figure:fig7} can be easily rescaled for different $\lambda_{L}$ values ranging from 0.36 to 1.0 for $\Delta M = 5$ GeV. These limits within the $M_{L}-\lambda_{L}$ plane are depicted in Figure \ref{figure:fig8}, where the shaded gray area delineated by the contour for $\Delta M = 5$ is excluded at the 95\% CL.

These predictions indicate that the mass of the VLL can be excluded in the range of 20 to 100 GeV, as long as the coupling constant \(\lambda_{L}\) is between 0.83 and 1.0. 
This exclusion is valid when the mass difference (\(\Delta M\)) is 5 GeV and with \(\mathcal{L_{\text{int}}} = 10.8\) ab\(^{-1}\) and \(\sqrt{s} = 240\) GeV. 

However, the FCC-ee might not possess the sensitivity needed to thoroughly investigate the VLL model, especially for cases where \(\lambda_{L}\) drops below 0.36, particularly with a small mass splitting of \(\Delta M = 5\) GeV.


\section{Summary}
\label{section:Summary}
The FCC-ee is a multipurpose $e^+ e^-$ collider which will be capable of scanning $\sqrt{s} $ ranging from 240 to 365 GeV. One of its aims is to discover physics beyond the SM.

In this analysis, we examined the production of lepton portal dark matter, specifically focusing on an electron-philic scalar dark matter scenario. In this scenario, dark matter particles are produced from the decay of pair-produced vector-like leptons, which are generated in $e^+ e^-$ collisions at the FCC-ee with $\sqrt{s} =$ 240 GeV. These vector-like leptons subsequently decay into a scalar dark matter particle and an electron. We simulated both signal and background events for the $e^+ e^-$ collisions at this energy, assuming $\mathcal{L_{\text{int}}} = 10.8$ ab$^{-1}$, which corresponds to FCC-ee Run 1. 

This analysis, which focused on the events meeting the final selection cuts, permits us to establish limits on the VLL model cross-sections. In addition, we determined exclusion limits on the mass of the doublet vector-like lepton (L) in scenarios with small mass differences between the VLL and the DM candidates, considering the mass splittings $\Delta M = 5$ GeV while varying the Yukawa coupling ($\lambda_L$).

In the case that \(\Delta M = 5\) GeV, the exclusion of $M_{L}$ from 20 up to 100 GeV has been established for a Yukawa coupling \(\lambda_{L}\) ranging from 0.83 to 1.0. This region could not be excluded by the LHC searches for vector-like leptons \cite{cms_vll, vll_atlas}.
These limits are done based on $\mathcal{L_{\text{int}}} = 10.8$ ab\(^{-1}\) at $\sqrt{s} =$ 240 GeV. However, if \(\lambda_{L}\) falls below 0.36, the FCC-ee is unlikely to have enough sensitivity to explore this VLL model thoroughly with the use of current parameters. 

This study highlights the exciting potential to explore situations where very light leptons (VLLs) are nearly degenerate in mass with dark matter (DM), especially when the mass difference, $\Delta M$, is significantly less than 100 GeV. It also aims to establish constraints on the Yukawa coupling, $\lambda_{L}$, at the FCC-ee.

\begin{acknowledgments}
The author of this paper would like to extend heartfelt thanks to J. Kawamura, the co-author of \cite{lpdm}, for providing the UFO model files and helping with the generation of signal events. Moreover, the research presented here has been made possible by the support of the Science, Technology, and Innovation Funding Authority (STDF) under grant number 48289.
\end{acknowledgments}



\begin{thebibliography}{9}

\bibitem{wimps} Scherrer, Robert J. and Turner, Michael S., On the relic, cosmic abundance of stable, weakly interacting massive particles, Phys. Rev. D 33 (1986) 1585.

\bibitem{planck2015} Planck Collaboration, Planck 2015 results. XIII. Cosmological parameters, Astron. Astrophys. 594 (2016) A13.

\bibitem{planck2018} Planck 2018 results. VI. Cosmological parameters. A\&A 641, A6 (2020).

\bibitem{Chang:2014tea}
S.~Chang, R.~Edezhath, J.~Hutchinson and M.~Luty, Leptophilic Effective WIMPs, Phys. Rev. D 90 (2014) no.1, 015011.

\bibitem{Bai:2014osa}
Y.~Bai and J.~Berger, Lepton Portal Dark Matter, JHEP 08 (2014), 153.

\bibitem{susy1}
S. P. Martin, Extra vectorlike matter and the lightest Higgs scalar boson mass in low-energy supersymmetry, Phys. Rev. D 81 (2010) 035004.

\bibitem{susy2}
S. Zheng, Minimal vectorlike model in supersymmetric unification, Eur. Phys. J. C 80 (2020) 273.

\bibitem{susy3}
M. Endo, K. Hamaguchi, S. Iwamoto, and N. Yokozaki, Higgs mass and muon anomalous magnetic moment in supersymmetric models with vectorlike matters, Phys. Rev. D 84 (2011) 075017.

\bibitem{gut1}
R. Nevzorov, E6 inspired supersymmetric models with exact custodial symmetry, Phys. Rev. D 87 (2013) 015029.

\bibitem{gut2}
I. Dorsner, S. Fajfer, and I. Mustac, Light vector-like fermions in a minimal SU(5) setup, Phys. Rev. D 89 (2014) 115004.

\bibitem{gut3}
A. Joglekar and J. L. Rosner, Searching for signatures of E6, Phys. Rev. D 96 (2017)015026.

\bibitem{extra_dim1}
G. Y. Huang, K. Kong, and S. C. Park, Bounds on the fermion-bulk masses in models with universal extra dimensions, JHEP 06 (2012) 099.

\bibitem{extra_dim2}
K. Kong, S. C. Park, and T. G. Rizzo, A vector-like fourth generation with a discrete symmetry from Split-UED, JHEP 07 (2010) 059.

\bibitem{vll_dm1}
P. Schwaller, T. M. P. Tait, and R. Vega-Morales, Dark matter and vectorlike leptons from gauged lepton number, Phys. Rev. D 88 (2013) 035001.

\bibitem{vll_dm2}
J. Halverson, N. Orlofsky, and A. Pierce, Vectorlike leptons as the tip of the dark matter iceberg, Phys. Rev. D 90 (2014) 015002.

\bibitem{vll_dm3}
S. Bahrami et al., Dark matter and collider studies in the left-right symmetric model with vectorlike leptons, Phys. Rev. D 95 (2017) 095024.

\bibitem{vll_dm4}
S. Bhattacharya, P. Ghosh, N. Sahoo, and N. Sahu, Mini review on vector-like leptonic dark matter, neutrino mass, and collider signatures, Front. Phys. 7 (2019) 80.

\bibitem{hierarchy1}
A. Falkowski, D. M. Straub, and A. Vicente, Vector-like leptons: Higgs decays and collider phenomenology, JHEP 05 (2014) 092.

\bibitem{hierarchy2}
K. Agashe, T. Okui, and R. Sundrum, Common origin for neutrino anarchy and charged hierarchies, Phys. Rev. Lett. 102 (2009) 101801.

\bibitem{hierarchy3}
M. Redi, Leptons in composite MFV, JHEP 09 (2013) 060.

\bibitem{muon_moment1}
Radovan Dermisek, Aditi Raval, Explanation of the muon g-2 anomaly with vectorlike leptons and its implications for Higgs decays, Phys. Rev. D 88 (2013) 013017.

\bibitem{muon_moment2}
Eugenio Megias, Mariano Quiros, and Lindber Salas, $g_{\mu} - 2$ from vector-like leptons in warped space, JHEP 05 (2017) 016.

\bibitem{muon_moment3}
J. Kawamura, S. Raby, and A. Trautner, Complete vectorlike fourth family and new $U^{\prime}(1)$ for muon anomalies, Phys. Rev. D 100 (2019) 055030.

\bibitem{muon_moment4}
G. Hiller, C. Hormigos-Feliu, D. F. Litim, and T. Steudtner, Model building from asymptotic safety with Higgs and flavor portals, Phys. Rev. D 102 (2020) 095023.

\bibitem{lpdm}
Junichiro Kawamura, Shohei Okawa, Yuji Omura, Current status and muon g-2 explanation of lepton portal dark matter, JHEP 08 (2020) 042.

\bibitem{Kawamura:2022uft}
J.~Kawamura, S.~Okawa and Y.~Omura, W boson mass and muon g-2 in a lepton portal dark matter model, Phys. Rev. D 106 (2022) no.1, 015005.

\bibitem{cms_vll}
CMS Collaboration, Review of searches for vector-like quarks, vector-like leptons, and heavy neutral leptons in proton-proton collisions at $\sqrt{s}$ = 13 TeV at the CMS experiment, Phys. Rept. 1115 (2025) 570.

\bibitem{vll_atlas}
ATLAS Collaboration, Search for vector-like leptons coupling to first- and second-generation Standard Model leptons in pp collisions at $\sqrt{s}$ = 13 TeV with the ATLAS detector, JHEP 05 (2025) 075.




\bibitem{FCC} I. Agapov et. al., Future Circular Lepton Collider FCC-ee: Overview and Status. DOI: https://doi.org/10.48550/arXiv.2203.08310.

\bibitem{CLIC} M. Aicheler et.al., The Compact Linear Collider (CLIC) - Project Implementation Plan, DOI: https://e-publishing.cern.ch/index.php/CYRM/issue/view/68.





\bibitem{whizard}
Wolfgang Kilian, Thorsten Ohl, and Jurgen Reuter, WHIZARD—simulating multi-particle processes at LHC and ILC. The European Physical Journal C, 71(9), Sep 2011.

\bibitem{Omega} T. Ohl. OMega  \& WHIZARD: Monte Carlo event generator generation for future colliders. DOI:10.1063/1.1394396.

\bibitem{pythia6}T. Sjostrand, S. Mrenna and P. Skands, PYTHIA 6.4 Physics and Manual, arXiv:hep-ph/0603175.

\bibitem{idea} 
The IDEA detector concept for FCC-ee (2025), arXiv:2502.21223 [physics.ins-det].

\bibitem{delphes}J. de Favereau, C. Delaere, P. Demin, A. Giammanco, V. Lematre, A. Mertens, M.Selvaggi, DELPHES 3, A modular framework for fast simulation of a generic collider experiment, JHEP 1402 (2014).


\bibitem{delphes-systematics}
Ang Li, Jan Eysermans, Gregorio Bernardi, Kevin Dewyspelaere, Michele Selvaggi, Christoph Paus, Model-independent ZH production cross section at FCC-ee. 	
arXiv:2512.21290 [hep-ex].

\bibitem{R58} A. L. Read, Presentation of search results: the CLs technique, J. Phys. G: Nucl. Part. Phys. 28 (2002) 2693.

\bibitem{R59} T. Junk, 
Confidence level computation for combining searches with small statistics,
Nuclear Instruments and Methods in Physics Research Section A: Accelerators, Spectrometers, Detectors and Associated Equipment,
Volume 434, Issues 2–3,
1999,
Pages 435-443,
ISSN 0168-9002.

\bibitem{R2} G. Cowan et al., 
Asymptotic formulae for likelihood-based tests of new physics,
Eur. Phys. J. C 71 (2011), p. 1554, Erratum: Eur. Phys. J. C 73 (2013) 2501.

\end{thebibliography}
\end{document}